\documentclass[twocolumn]{emulateapj}

\usepackage{graphicx}
\usepackage{natbib}
\usepackage{multirow}
\usepackage{amsmath}
\usepackage{amssymb}
\usepackage{rotating}
\usepackage{hyperref}
\usepackage{xspace}

\newcommand{\hii}{\ion{H}{2}\xspace}
\newcommand{\am}{NH$_{3}$\xspace}
\newcommand{\amiso}{$^{15}$NH$_{3}$\xspace}
\newcommand{\ammain}{$^{14}$NH$_{3}$\xspace}

\newcommand{\kms}{km s$^{-1}$\xspace}

\begin{document}



\title{Discovery of \ammain\, (2,2) maser emission in Sgr B2-Main}

\author{E. A. C. Mills}
\affil{Physics Department, Brandeis University, 415 South Street, Waltham, MA 02453, USA}
\email{elisabeth.ac.mills@gmail.com}

\author{ A. Ginsburg}
\affiliation{National Radio Astronomy Observatory\altaffilmark{2} 1003 Lopezville Rd Socorro, NM 87801, USA}

\author{A. R. Clements}
\affil{Department of Chemistry, University of Virginia, Charlottesville, VA, 22904, USA}

\altaffiltext{1}{A. Ginsburg is a Jansky Fellow of the National Radio Astronomy Observatory.}
\altaffiltext{2}{The National Radio Astronomy Observatory is a facility of the National Science Foundation operated under cooperative agreement by Associated Universities, Inc.}

\author{P. Schilke}
\affiliation{I. Physikalisches Institut, Universit\"{a}t zu K\"{o}ln, Z\"{u}lpicher Str. 77, D-50937 K\"{o}ln, Germany}

\author{\'{A}. S\'{a}nchez-Monge}
\affiliation{I. Physikalisches Institut, Universit\"{a}t zu K\"{o}ln, Z\"{u}lpicher Str. 77, D-50937 K\"{o}ln, Germany}

\author{K. M. Menten}
\affiliation{Max-Planck-Institut f\"{u}r Radioastronomie, Auf dem H\"{u}gel 69, 53121 Bonn, Germany}

\author{N. Butterfield}
\affil{Green Bank Observatory, 155 Observatory Rd, PO Box 2, Green Bank, WV 24944, USA  }

\author{C. Goddi}
\affil{Department of Astrophysics/IMAPP, Radboud University, PO Box 9010, 6500 GL Nijmegen, The Netherlands}

\author{A. Schmiedeke}
\affil{Max-Planck Institute for Extraterrestrial Physics, Giessenbachstrasse 1, D-85748 Garching, Germany}

\author{C. G. De Pree}
\affil{Department of Physics \& Astronomy, Agnes Scott College, 141 E. College Ave., Decatur, GA 30030, USA}

\begin{abstract}
We report the discovery of the first \ammain\, (2,2) maser, seen in the Sgr B2 Main star forming region near the center of the Milky Way, using data from the Very Large Array radio telescope. The maser is seen in both lower resolution (3$\arcsec$ or $\sim$0.1 pc) data from 2012 and higher resolution ($0\arcsec.1$ or $\sim$1000 AU) data from 2018. In the higher resolution data \am\, (2,2) maser emission is detected toward 5 independent spots. The maser spots are not spatially or kinematically coincident with any other masers in this region, or with the peaks of the radio continuum emission from the numerous ultracompact and hypercompact \hii\, regions in this area. While the (2,2) maser spots are spatially unresolved in our highest resolution observations, they have unusually broad linewidths of several kilometers per second, which suggests that each of these spots consists of multiple masers tracing unresolved velocity structure. No other \am\, lines observed in Sgr B2 Main are seen to be masers, which continues to challenge existing models of \am\, maser emission.

\end{abstract}

\section{Introduction}
Ammonia masers are relatively rare, and have been seen in only a handful of sources, mainly high-mass star forming regions. Despite this rarity, many different \ammain\, ($J,K$) inversion transitions, both nonmetastable ($J\neq K$) and metastable ($J=K$), have been observed to be masers \citep[e.g.,][]{Wilson93}. Maser action has even been reported to occur in a transition of the rare isotopologue \amiso\, \citep{Mauers86,Johnston89,Gaume91,Schilke91}. 

\ammain\, masers arising from nonmetastable energy levels are more frequently seen than metastable masers. This is likely due to the smaller population of these levels, which quickly decay unless excited by extreme densities or intense radiation fields and are thus easier to invert \citep{Wilson93}. The most common of these appear to be in the ortho-\am\, ($K=3n$) transitions (9,6) and (6,3) \citep{Madden86,Pratap91,Wilson93}. However, few surveys for nonmetastable \ammain\, masers are reported in the literature (\citealt{Madden86} report surveying 17 regions for maser emission but only finding masers in 4 of these), and ultimately these masers have only been reported toward 5 regions: W51, W49, DR21, NGC 7538, and NGC 6334. Additional masing nonmetastable transitions, including para-\am\, transitions, have been observed in W51 \citep{Mauers87,Wilson88,Wilson90,Henkel13}, NGC 7538 \citep{Hoffman11,Hoffman14}, and NGC 6334 \citep{Walsh07}. 

Of the metastable \ammain\, masers, the most commonly observed masing transition is the ortho (3,3) line, which has been definitively seen toward DR21(OH), W51-North, and G5.89-0.39 \citep{Mangum94,Zhang95,Hunter08}, and which has been suggested to also be a maser in a variety of other sources, including NGC 6334, the southern part of Sgr B2, and the nucleus of the starburst galaxy NGC 253 \citep{Kraemer95,JMP99,Ott05}. Other ortho lines -- (6,6), (9,9), and (12,12)--  have also been reported to exhibit maser action in NGC 6334 \citep{Walsh07} and W51 \citep{Henkel13,Goddi15}. Para-\am\, masers are rarer, however masers have been reported in the (1,1) line toward the DR21 \hii\, region \citep{Gaume96}, in the (5,5) line toward G9.62+0.19 \citep{Hofner94} and in the (7,7) line toward W51 North\citep{Goddi15}.

We report the discovery of a new, strong para-\ammain\ maser in the metastable (2,2) transition toward the Galactic center star forming region Sgr B2 Main, hereafter Sgr B2 (M). Sgr B2 is the most massive molecular cloud in the central 500 parsecs of the Galaxy, accounting for 10\% of the mass in this region \citep{Scoville1975}. This cloud hosts the most intense ongoing star formation in the Galactic center, much of which is concentrated in two sites of clustered star formation: Sgr B2 (M) and Sgr B2 North or (N). Sgr B2 (M) is suggested to be older than Sgr B2 (N), as it is less chemically rich \citep{Corby15,SM17} and appears more fragmented \citep{Qin11,SM17,Ginsburg18}. In addition to containing a fragmented,  massive hot core \citep{Vogel87,Qin11,SM17}, it contains dozens of ultracompact and hypercompact \hii\, regions \citep{dePree98}. Sgr B2 (M) also contains at least one molecular outflow traced by H$_2$O masers, \ammain\, emission and absorption, and millimeter H$_2$CO lines \citep{Vogel87,McGrath04,Qin08}. In this paper, we present new, high-resolution centimeter-wave observations of \ammain\, toward this region, and discuss the detected (2,2) maser emission both in the context of other star formation indicators detected in this specific region, and in the context of \am\, masers in general.

\section{Data}
\subsection{Low-resolution observations}
Observations of \ammain\, toward Sgr B2 (M) were made using the Karl G. Jansky Very Large Array (VLA), a facility of the National Radio Astronomy Observatory\footnotemark[1]\footnotetext[1]{The National Radio Astronomy Observatory is a facility of the National Science Foundation operated under cooperative agreement by Associated Universities, Inc.} in January 2012 in DnC configuration using the K band receivers (project 11B-210, PI: E.A.C. Mills). For the \ammain\, (2,2) line at 23.7226333 GHz, the synthesized beam size was 2\arcsec.83 $\times$ 2\arcsec.56, which corresponds to a spatial resolution of 0.1 pc at the adopted distance to Sgr B2 of 8 kpc \citep{Reid09,Gravity18}. The velocity resolution of the (2,2) line observations was 1.58 \kms. In addition to the (2,2) line, the (1,1) through (7,7) and (9,9) metastable \ammain\, lines were also observed, along with the (10,9) nonmetastable line. The observations toward the Sgr B2 cloud were part of a larger survey of Galactic center clouds; additional description of these observations, including details on the data calibration and imaging, is given in \cite{Mills15}.

\subsection{High-resolution observations}
Observations of \ammain\, toward Sgr B2 (M) were also made using the A configuration of the VLA in March-April 2018 (project 18A-229, PI: A. Ginsburg). These observations had a spatial resolution of 0$''$.17$\times$0$''$.08 (1360$\times$640 AU) and a velocity resolution of 0.79 \kms\, for the (2,2) line. Other observed lines of \ammain\, included (1,1), (2,2), (4,4), (5,5), (7,7), (2,1), (3,2), (5,3), and (5,4), however the (3,2) line was not observed in all sources as it was on the edge of a sub-band which limited the frequency coverage. 

The data were pipeline processed using the VLA pipeline after removing the automated RFI flagging step.  We then performed three iterations of phase-only self-calibration using the brightest \ammain\, (2,2) maser peak.

\subsection{Absolute positional reference frame}
We define the absolute positional reference frame for these observations using a high-resolution (0$\arcsec$.2), 3 mm ALMA continuum map (Project code 2016.1.00550.S, PI: A. Ginsburg) which should have good absolute positional accuracy as it is tied to a larger mosaic of the region \citep{Ginsburg18}. We use the task `register\_translation' from the python package scikit-image \citep{scikit14} to perform cross correlations between the ALMA image and our A-configuration continuum image, and shift our A-configuration images by $0\arcsec.05$ in Right Ascension and $0\arcsec.48$ in Declination. The uncertainty in the alignment from the cross correlation is $\pm0\arcsec.03$. This large offset is due to an error in the VLA calibrator catalog for the position of source J1744-3116, which is used as a phase calibrator for the A-configuration data (L. Sjouwerman, private comm.).


\begin{figure}
\includegraphics[scale=0.4]{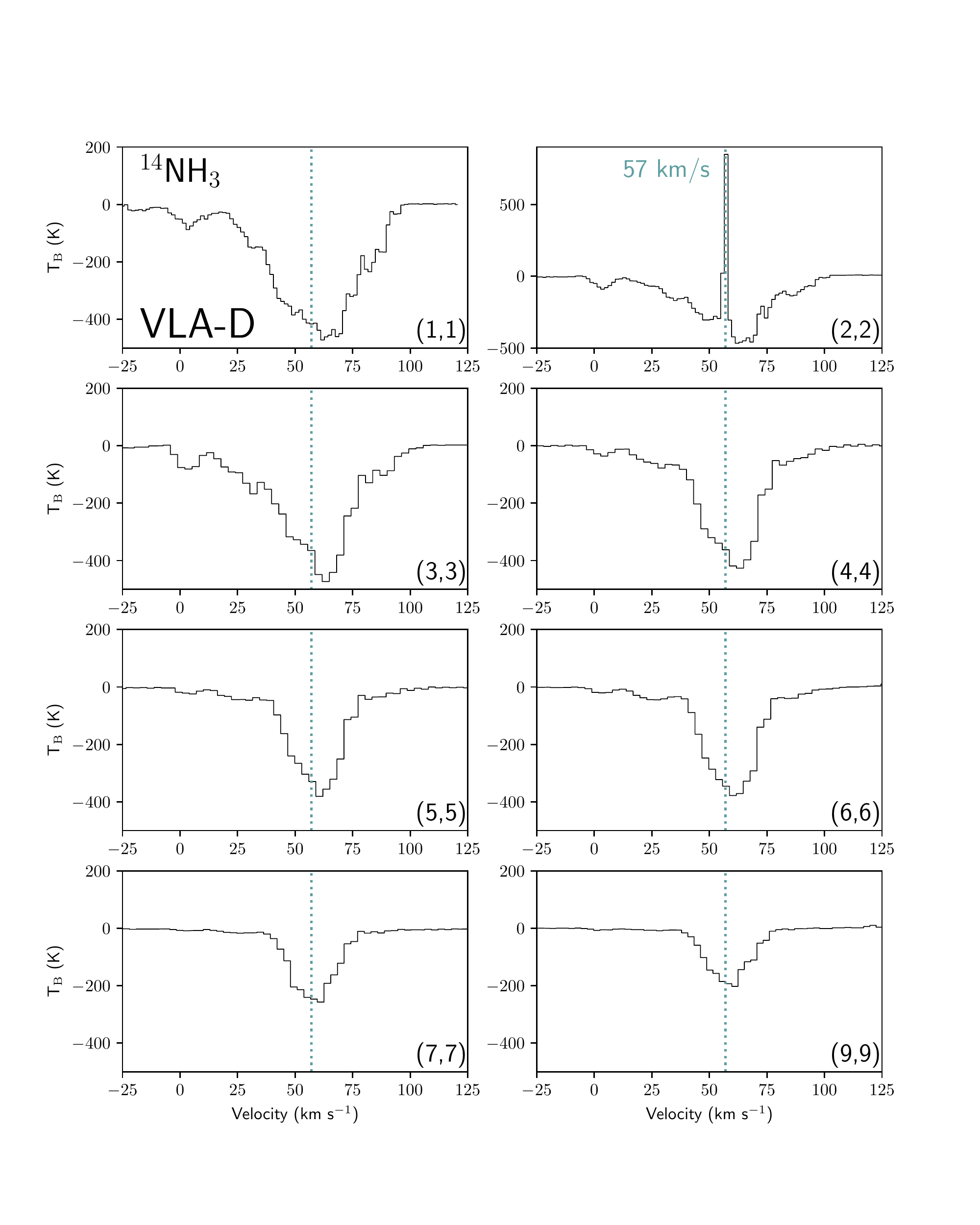}
\caption{Spectra of \ammain\, metastable transitions from VLA D-configuration observations. All transitions are seen in absorption against the `F' complex of \hii\, regions in Sgr B2 (M) except the (2,2) line, where maser emission is also seen at a velocity of 57 \kms. This velocity is indicated by a vertical dotted line in each panel.}
\label{Dconfig}
\end{figure}

\section{Results}

\begin{figure}
\includegraphics[scale=0.4]{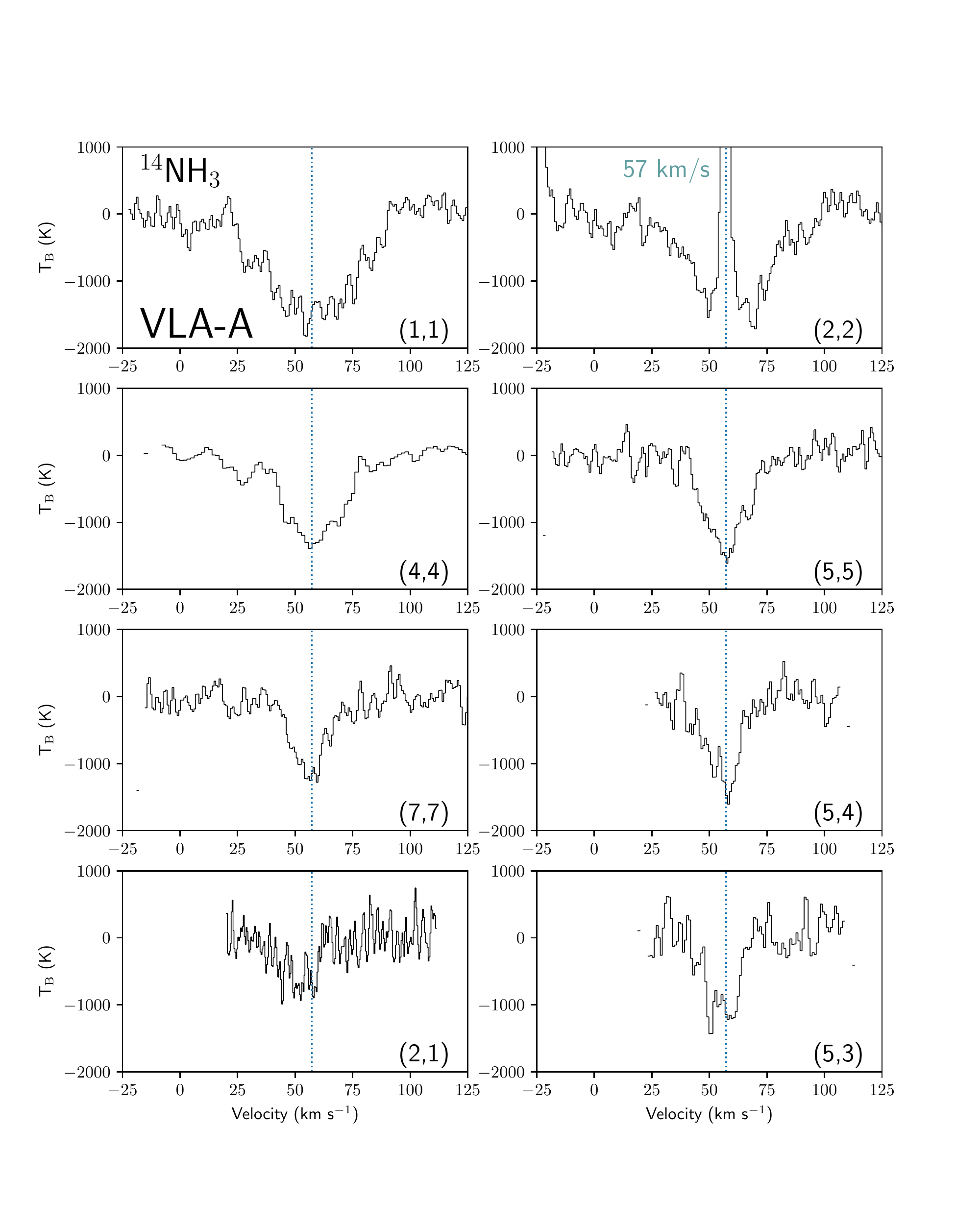}
\caption{Spectra of \ammain\, metastable and nonmetastable transitions from VLA A-configuration observations toward the position of Maser M1, the strongest (2,2) maser spot. All transitions are seen in absorption against the `F' complex of \hii\, regions in Sgr B2 (M) except the (2,2) line, where the maser emission is seen at a central velocity of $\sim$57 \kms (vertical dotted line). The y-axis of all of the plots has been scaled to the same range to emphasize the absorption present in all transitions; the peak of the (2,2) maser emission is 9$\times10^4$ K.}
\label{Aconfig}
\end{figure}

\begin{figure}
\includegraphics[scale=0.4]{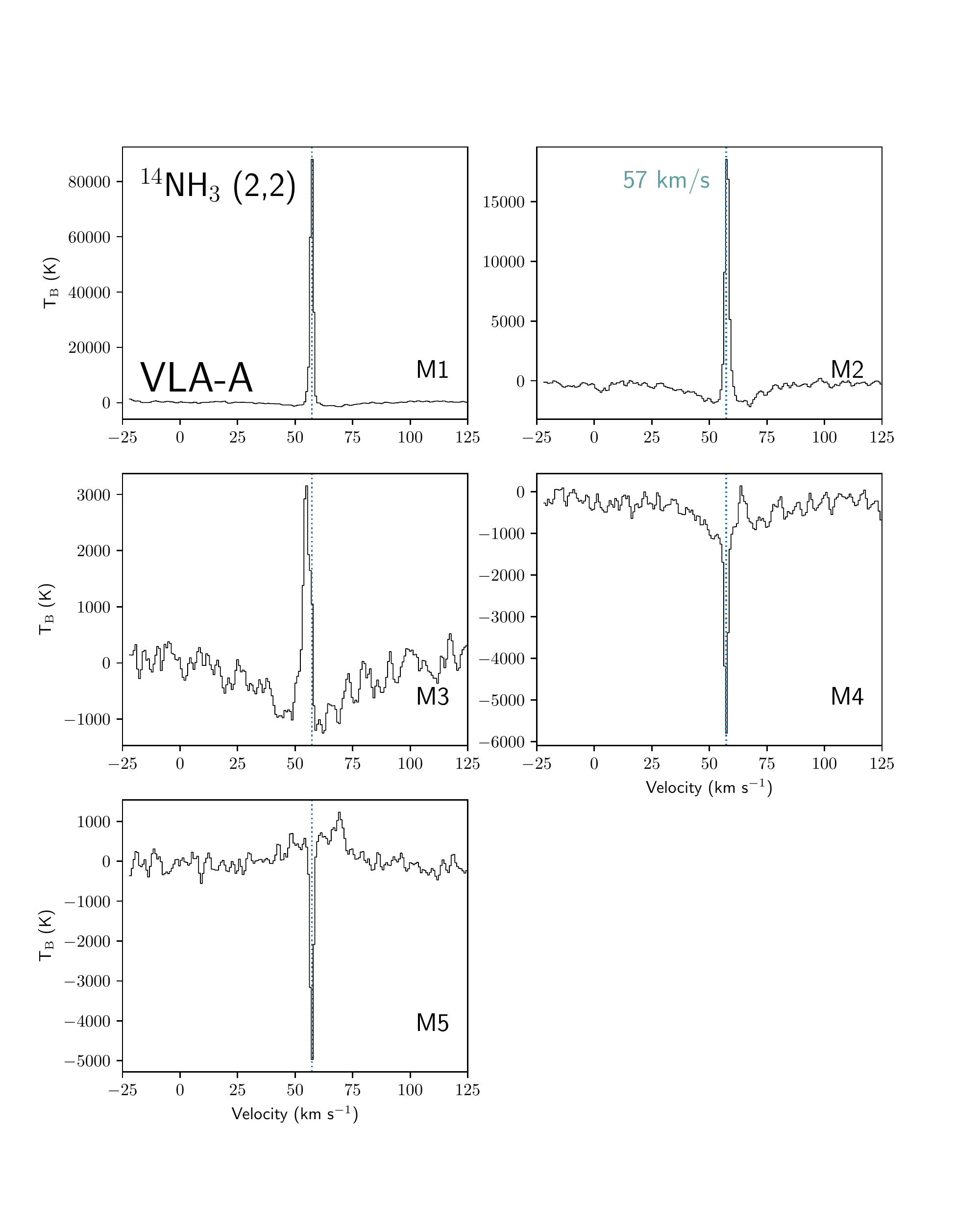}
\caption{Spectra of the \ammain\, (2,2) line toward each of the maser spots seen in the A-configuration data.  Maser spots M1-4 are superimposed on absorption against nearby \hii\, regions, while M5 lies on top of a broad, thermal emission profile. The central velocity of M1, the brightest maser source, is shown as a vertical dotted line at 57 \kms\, in each panel. The narrow absorption at this velocity seen in the spectra of M4 and M5 is an artifact from the sidelobes of M1.}
\label{all_masers}
\end{figure}

Of the 13 observed transitions of \ammain\, (10 of which were observed in A configuration), we detect maser emission in only the (2,2) line\footnotemark[1]\footnotetext[1]{We investigate other possible line identifications for this maser, and find that there only three species that have been detected toward Sgr B2 (M) that have transitions lying within $\sim$10 km s$^{-1}$ of the (2,2) line: HCOOCH$_3$ (and its isotopologue H$^{13}$COOCH$_3$ which has not been detected but is likely present), $^{17}$OH and CH$_3$OCH$_3$. However, none of these are a plausible candidate for this maser, as these species have either never been observed to be masers in the ISM ($^{17}$OH and CH$_3$OCH$_3$) or have only been observed to be weak masers with widths similar to those of thermally excited lines \citep{Faure14}, which is inconsistent with the narrower profiles seen in our observations.}. The A-configuration observations of \ammain\, (5,4) also overlap with the \amiso\, (2,2) line, which we can confirm does not show similar maser emission. 
In Figure \ref{Dconfig} we show spectra from the D-configuration observations in which we see a single, narrow (the line is unresolved for a channel width of 1.5 \kms) maser component at 57 \kms, in the midst of broad absorption that is seen in all of the observed \ammain\, lines. In the A configuration, this maser breaks up into 5 independent spots spread out over 2$''$ at velocities from 57 to 69 \kms, with measured velocity widths of 1.6-3.9 \kms. We show spectra of all the \ammain\, lines toward the brightest maser spot in Figure \ref{Aconfig}, and profiles of the (2,2) line toward each of the 5 detected maser spots in Figure \ref{all_masers}. The (2,2) maser appears to be a persistent feature over a large time baseline, as it is not only seen in our 2012 and 2018 observations, but can be seen in Figure 6 of \cite{Vogel87}, where there is a negative optical depth (i.e., emission) at a velocity of $\sim$ 60 \kms. 

\begin{figure}
\includegraphics[scale=0.32]{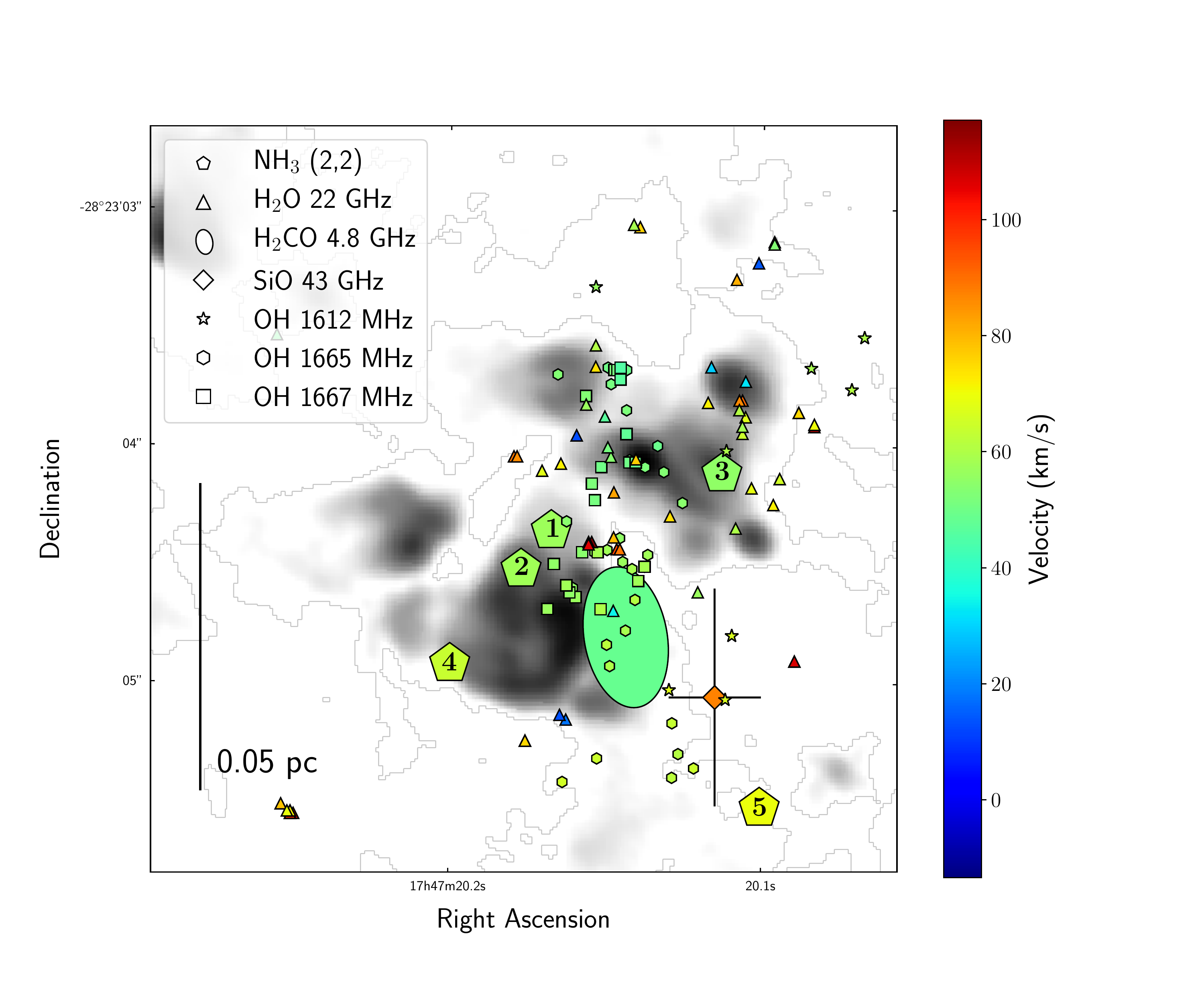}
\caption{Position of the \ammain\, (2,2) masers detected toward Sgr B2 (M), overlaid on a 3 mm ALMA continuum map (Project 2016.1.00550.S, PI: A. Ginsburg). Also shown are the positions of published maser detections from H$_2$O \citep{McGrath04}, OH \citep{Gaume90}, SiO \citep{Morita92}, and H$_2$CO \citep{Mehringer94}.  Only the OH masers brighter than 1 Jy are shown, due to the large number detected in this region. The positions of all masers except SiO are determined relative to the continuum emission, and should have positional uncertainty $<0\arcsec.5$, smaller than the plotted points. The H$_2$CO maser emitting region is partially resolved and plotted as an ellipse.} 
\label{maser}
\end{figure}

The positions of the \ammain\, masers are shown in Figure \ref{maser}. The brightest maser is M1, which has a peak intensity of 0.5~Jy, or a brightness temperature of T$_B>9\times 10^4$ K.\footnotemark[2]\footnotetext[2]{All listed brightness temperatures are lower limits, as the maser spots are spatially unresolved.} M2 is only slightly offset from M1, and its spectrum contains an additional emission contribution from the sidelobes of M1; it may peak nearer to a velocity of 59 \kms\, with T$_B>7000$ K. M3 is offset by $\sim1\arcsec$ from M1 and has T$_B>$3900 K. M4 and M5 are much weaker, having limits on T$_B$ of only $>700 K$. While this could be considered consistent with thermal emission, as temperatures $>1000$ K have been measured toward Sgr B2 (M) \citep{Wilson06}, no correspondingly bright emission is seen in the other \am\, lines, indicating that these sources are also nonthermal.  Maser spots M1-M4 are on the edge of continuum sources, and the maser emission toward these regions is superimposed on a broad absorption feature. However, M5 is not near to any known continuum source, and shows a weak thermal emission rather than absorption profile in the other \am\, lines. In Table \ref{maser_stats} we list the properties of each detected maser spot, including its position, central velocity, velocity width, and peak brightness temperature.

\begin{table*}[t]
\caption{Measured Maser Spot Parameters}
\centering
\begin{tabular}{l l l l l l}
\hline\hline
Source & Right Ascension & Declination & v$_{\mathrm{cen}}$ & v$_{\mathrm{fwhm}}$ & Peak T$_{\mathrm{MB}}$ \\
       & & & (km s$^{-1}$)      & (km s$^{-1}$)       & (K)  \\
\hline\hline
M1 & 17$^{\mathrm h}$47$^{\mathrm m}$20$^s$.167$\pm$0.001 & -28$\degr$23$'$4$\arcsec$.36$\pm$0.01 &  57.22 $\pm$ 0.01 & 1.64 $\pm$ 0.02 & 90900 $\pm$ 1000 \\
M2 & 17$^{\mathrm h}$47$^{\mathrm m}$20$^s$.177$\pm$0.001 & -28$\degr$23$'$4$\arcsec$.52$\pm$0.02 & 57.68 $\pm$ 0.04 & 2.13 $\pm$ 0.10 & 21100 $\pm$ 800 \\
M3 & 17$^{\mathrm h}$47$^{\mathrm m}$20$^s$.113$\pm$0.002 & -28$\degr$23$'$4$\arcsec$.11$\pm$0.04 & 54.99 $\pm$ 0.20 & 3.89 $\pm$ 0.47 & 3900 $\pm$ 400 \\
M4 & 17$^{\mathrm h}$47$^{\mathrm m}$20$^s$.200$\pm$0.001 & -28$\degr$23$'$4$\arcsec$.92$\pm$0.02 & 64.05 $\pm$ 0.51 & 2.15 $\pm$ 1.19 & 700 $\pm$ 300 \\
M5 & 17$^{\mathrm h}$47$^{\mathrm m}$20$^s$.101$\pm$0.001 & -28$\degr$23$'$5$\arcsec$.52$\pm$0.02 & 69.21 $\pm$ 0.46 & 3.43 $\pm$ 1.09 & 700 $\pm$ 200 \\
\hline\hline
\end{tabular}
\label{maser_stats}
\end{table*}

Based on the correspondence between the absorption and emission seen in the A-configuration observations, we believe it is most likely that the main hyperfine component\footnotemark[1]\footnotetext[1]{The inner and outer hyperfine satellites in the (2,2) line are offset from the main hyperfine line component by $\pm$16.6 and $\pm$25.8 \kms, respectively.} of the (2,2) line  is masing \citep[while this is assumed for most observed \ammain\, masers, the only observed 1,1 maser occurs in an outer hyperfine satellite component;][]{Gaume96}. The maser velocity is also consistent with the velocity of relatively narrow metastable line emission seen in the D-configuration observations 4$''$ to the west. However, the maser velocity is not consistent with the velocity of \ammain\, metastable and nonmetastable line emission observed 3$''$ to the south, which appears to be closer to 70 \kms. 

Comparing the maser positions to the radio continuum in Figure \ref{maser}, we do not see a clear correspondence between these masers and radio continuum: only four of the five masers overlap with continuum emission, and these tend to be located on the edge of the \hii regions in Sgr B2 (M). Interestingly, M1 is located within $0\arcsec.2$ of an increase in continuum emission from source `F3c' detected in \cite{dePree14}. We also compare the position of the (2,2) masers to a number of other masers, both rare and common, that are also seen in Sgr B2 (M). This includes 1612, 1665, and 1667 MHz OH masers \citep{Gaume90}, 22 GHz H$_2$O masers \citep{McGrath04}, a 4.8 GHz H$_2$CO maser \citep{Mehringer94}, and a 43 GHz $J=1-0, v=1$ SiO maser \citep{Morita92}. We have used published figures of the masers and the simultaneously observed continuum emission to align the maser positions with our continuum maps. We adjust the positions of the H$_2$O masers by $+0\arcsec.16$ in Right Ascension and $+0\arcsec.54$ in Declination to match our reference frame. We also adjust the position of the OH masers by $+0\arcsec.84$ in Declination and the H$_2$CO maser by $+0\arcsec.42$ in Declination. No corresponding continuum map for the 43 GHz SiO maser is published in \citep{Morita92}, so we adopt their published position and estimates of the absolute positional uncertainty.  The position of these other masers relative to the (2,2) masers is shown in Figure \ref{maser}. None of the other masers observed  in Sgr B2 (M) appear to overlap with the (2,2) maser in both position and velocity, though there is a weak ($<2Jy$) OH 1665 MHz maser within $0\arcsec.1$ of M1 and offset by $\sim$ 5 \kms. 

\section{Discussion}

\subsection{Interpreting the (2,2) maser in Sgr B2 (M)}

This is the first \ammain (2,2) maser seen in any source. It is a relatively strong maser, with brightness temperatures constrained to be $>10^4$ K by the A-configuration observations. This is among the brightest \ammain\, masers that have been reported, which typically have limits on their brightness temperatures of $>10^2-10^5$ K \citep{Madden86,Wilson91,Mangum94,Hofner94,Zhang95,Gaume96,Hunter08,Walsh07,Hoffman11,Hoffman14,Goddi15}, with only a single VLBI measurement of the \ammain\, (9,6) maser having a brightness temperature $>10^{13}$ K\citep{Pratap91}. 

The most unique characteristic of the observed (2,2) masers is their linewidth. In the A-configuration observations, line profiles of individual maser spots are well resolved, having widths of up to 4 \kms. Narrow ($<$1\kms) or unresolved widths are typical of most other observed \am\, masers. 
As individual masers show nonthermal linewidths, we must assume multiple masers are present, and that the extreme or unique physical conditions that lead to these masers are relatively common in this region. Given that both the (2,2) masers and the nearby SiO maser show relatively broad linewidths over a compact region, an intriguing possibility is that these masers could arise in or near rotating protostellar disks \citep[e.g., as in the SiO masers around Orion's source I;][]{Goddi09,Matthews10}

Given limits on the observed maser spot sizes of $\lesssim$1000 AU, the observed (2,2) maser line widths  could be consistent with Keplerian rotation in the disk of a low- or intermediate-mass protostar. A line width of 2-4 \kms for a disk with a radius of 400 AU would be consistent with rotation around a central mass of 0.5-2 M$_\odot$. In the future, one might be able to test whether the \ammain\, masers arise from such organized or compact circumstellar structures by making VLBI observations of these masers. However, such low mass protostellar sources would be difficult to detect in millimeter continuum emission as imaging near the bright sources in this region is significantly limited by an achievable dynamic range of $\sim$5000 with ALMA \citep{Ginsburg18}.

\subsection{The Excitation of the (2,2) Maser}

Our observations suggest that maser emission is possible in essentially any metastable transition of \ammain: of the inversion transitions up to $J=9$, only the (4,4) and (8,8) lines have not yet been observed to show maser action, though a weak \amiso\, (4,4) maser has been reported \citep{Schilke91}. However at present, no universally successful mechanism has been proposed for \am\, masers.  Suggested models for \am\, maser excitation involve three possible mechanisms (1) radiative excitation from 10 $\mu$m continuum emission, (2) collisional excitation, and (3) radiative excitation from a chance line overlap.

Current models of nonmetastable maser emission favor mechanism (1): excitation through the 10 $\mu$m vibrational transitions pumped by infrared emission \citep{Brown91}. However, these theories tend to predict that multiple, adjacent transitions of \am\, should mase simultaneously \citep{Wilson93}. While masers have been seen in adjacent nonmetastable transitions \citep{Henkel13,Hoffman14}, and metastable transitions of ortho-\am\, \citep[e.g., 6,6 and 9,9 masers;][]{Goddi15}, no sources have been observed to show maser emission in multiple para-\am\, transitions. The only published model of a metastable \am\, involves the second mechanism, pumping the ortho (3,3) transition via collisions with H$_2$ \citep{Flower90,Mangum94}. However, this model does not predict masers in metastable transitions of para \am\, \citep{Schilke91}, and cannot explain observations of simultaneous maser emission in the ortho- (4,3) and (3,3) lines of \amiso\, \citep[][]{Schilke91}. Mechanism (3), line overlap, is not currently favored to explain any of the observed \am\, masers \citep[e.g., as discussed in ][]{Goddi15}. Indeed, the detection of maser emission in the (2,2) line makes this more unlikely as a general mechanism to explain the metastable para-\am\, masers, as separate instances of line overlap would be needed to pump each of the (1,1), (2,2), (5,5), and (7,7) lines. 

 As none of the existing three simple models provide a satisfactory explanation for these masers, we suggest that the excitation of the metastable para-\am\, masers is likely complicated, and may require a confluence of geometry, velocity structure and possibly both collisional and radiative excitation (rotational and rovibrational transitions  of \am\, occur at a range of infrared wavelengths from 9 $\mu$m to 400 $\mu$m, and the vibrational transitions in particular may be influenced by the shape of the overlapping 9.7 $\mu$m silicate absorption feature, as noted by \citealt{Barentine12}). Devising and validating such a complex model will likely require the availability of collisional coefficients for states above $J=6$, which is the limit of currently published data \citep[e.g.,][]{Bouhafs17}.

\subsection{The (2,2) Maser as a Probe of Star Forming Regions}

As \ammain\, is a common probe of gas in star forming regions, particularly the (1,1) and (2,2) lines, it is somewhat surprising that masers in this line have not previously been seen. \cite{Lu14} surveyed 62 high-mass star forming regions in the \ammain\, (1,1) and (2,2) lines with subparsec spatial resolution using the VLA's DnC configuration and did not report any (2,2) maser emission. With an 800 K brightness temperature for this maser in our DnC-configuration observations, similar masers in other star forming regions should have been readily apparent. It appears likely then that, like other metastable \ammain\, masers, this maser is quite rare, which would limit its general usefulness as a star formation probe.   

The (2,2) maser is the first \am\, maser confirmed to exist in Sgr B2. This increases the number of star forming regions known to host either metastable or nonmetastable \am\, masers from 7 to 8. We suggest that more massive star forming regions should be carefully surveyed for \am\, masers in a wide variety of nonmetastable and metastable transitions to better understand their overall incidence and association with various stages and structures of star formation, and to provide improved constraints for models of \am\, maser excitation. 
\section{Summary}

We report the discovery of the first \ammain\, (2,2) maser, which is detected in the Sgr B2 Main star forming region. Below, we summarize our main findings.

\begin{itemize}
\item{At the high resolution of our VLA A-configuration observations ($0\arcsec.$1 or $\sim$1000 AU) the (2,2) maser breaks into 5 independent spots with $T_B$ from 700-$9\times10^4$ K. These \ammain\, masers are not spatially or kinematically coincident with any other masers in this region, or with the peaks of the observed radio continuum emission.}
\item{The \ammain\, (2,2) masers are spatially unresolved in the A-configuration observations but have unusually broad linewidths of 1.5-4 \kms, which could trace the kinematics of circumstellar gas.}
\item{The (2,2) maser is the only \ammain\, maser seen in this region and the fourth metastable transition of para-\ammain\, observed to be a maser, which increases the inconsistencies between the observed \am\, masers and the existing models for this emission.}
\end{itemize}

\end{document}